\documentclass[3p,times,procedia]{elsarticle}
\usepackage{nupha_ecrc}

\volume{00}

\firstpage{1}

\journalname{Nuclear Physics A}

\runauth{}

\jid{nupha}

\jnltitlelogo{Nuclear Physics A}

\usepackage{amssymb}

\usepackage[figuresright]{rotating}

\newcommand{\pt}{{\mathbf{p}_T}}

\newcommand{\abspt}{{p_T}}

\newcommand{\dd}{\mathrm{d}}

\renewcommand*{\appendixname}{Appendix}

\begin{document}

\begin{frontmatter}



\dochead{XXVIIth International Conference on Ultrarelativistic Nucleus-Nucleus Collisions\\ (Quark Matter 2018)}

\title{Importance of initial and final state effects for azimuthal correlations in p+Pb collisions}


\author[label1]{Moritz Greif}
\author[label1]{Carsten Greiner}
\address[label1]{Institut f\"ur Theoretische Physik, Johann Wolfgang Goethe-Universit\"at,
	Max-von-Laue-Str.\ 1, D-60438 Frankfurt am Main, Germany}

\author[label2]{Bj\"orn Schenke}
\address[label2]{Physics Department, Brookhaven National Laboratory, Upton, NY 11973, USA}

\author[label3]{S\"oren Schlichting}
\address[label3]{Department of Physics, University of Washington, Seattle, WA 98195-1560, USA}

\author[label4]{Zhe Xu}
\address[label4]{Department of Physics, Tsinghua University and Collaborative Innovation Center of Quantum Matter, Beijing 100084, China}

\begin{abstract}
We study the influence and interplay of initial state and final state effects in the dynamics of small systems, focusing on azimuthal correlations at different multiplicities. To this end we introduce a new model, matching the classical Yang-Mills dynamics of pre-equilibrium gluon fields (IP-GLASMA) to a perturbative QCD based parton cascade for the final state evolution (BAMPS) on an event-by-event basis. Depending on multiplicity of the event, we see transverse momentum dependent signatures of the initial, but also the final state in azimuthal correlation observables, such as $v_2\left\lbrace2PC\right\rbrace(p_T)$.
In low-multiplicity events, initial state correlations dominate for transverse momenta $p_T>2~{\rm GeV}$, whereas in high-multiplicity events and at low momenta final state interactions dominate and initial state correlations strongly affect $v_2\left\lbrace2PC\right\rbrace(p_T)$ for $p_T>2~{\rm GeV}$ as well as the $p_T$ integrated $v_2\left\lbrace2PC\right\rbrace$. Nearly half of the final pT integrated $v_2\left\lbrace2PC\right\rbrace$ is contributed by the initial state in low-multiplicity events, whereas in high-multiplicity the share is much less. Based on Ref.~\cite{Greif:2017bnr}, we are now able to carry out a systematic multiplicity scan, probing the dynamics on the border of initial state dominated to final state dominated - but not yet hydrodynamic – regime. 
\end{abstract}


\end{frontmatter}


\section{Introduction}
\label{sec:Intro}
The measured azimuthal momentum anisotropies of produced particles in heavy ion collisions are well described in the framework of event-by-event hydrodynamics~\cite{Heinz:2013th}. 
Measurements in smaller collision systems such as p+p and p+A, in particular those of anisotropies in multi-particle correlation functions, have shown very similar features as those in heavy ion collisions. While calculations within the hydrodynamic framework have been quite
successful in describing observables in these small collision systems (even though the applicability of hydrodynamics becomes increasingly doubtful as the system size decreases and gradients increase~\cite{Romatschke:2016hle,Mantysaari:2017cni}),
alternative explanations relying entirely on intrinsic momentum
correlations of the produced particles can also reproduce many features
of the experimental data~\cite{Dusling:2015gta,Schlichting:2016sqo}. 
So far all calculations of multi-particle correlations in small collision systems have studied either only intrinsic momentum correlations or purely final state driven effects. 
Here we  present the first study where both effects are combined into a single framework to assess their relative importance~\cite{Greif:2017bnr}. 

\section{Intial and final state interactions}
We present a new model, where we compute initial state gluon Wigner-distributions from the Impact Parameter dependent Glasma model (IP-Glasma) \cite{Schenke:2012wb,Schenke:2012hg} and via sampling of individual gluons feed them into the partonic transport simulation 'Boltzmann approach to multiparton scatterings' (BAMPS)~\cite{Xu:2004mz}. 

We calculate the solution to the classical Yang-Mills equations of motion up to $\tau_0=0.2\,{\rm fm}/c$ following the standard procedures described in~\cite{Schenke:2012wb,Schenke:2012hg}, including event-by-event fluctuations of the proton's geometrical structure~\cite{Mantysaari:2016ykx}.
Event-by-event we extract fluctuating ensembles of individual gluons in a boost-invariant geometry. Their distribution is anisotropic in momentum space~\cite{Schenke:2015aqa,Schenke:2016lrs}, thus contains the intrinsic momentum space correlations of the color glass condensate picture. The classical Yang-Mills evolution includes rescattering effects at early times but the semiclassical description of the dynamics becomes inapplicable after a relatively short time when quantum effects become important and the subsequent dynamics is more appropriately described in terms of weakly interacting quasi-particles \cite{Baier:2000sb,Berges:2013fga}. This is why we simulate the dynamics within $0.2~\rm{fm/c} < \tau < 2.0~\rm{fm/c}$, with a 3+1-dimensional Boltzmann approach to multi-parton scatterings (BAMPS), which, starting from the initial phase-space density of the sampled gluons, solves the relativistic Boltzmann equation $p^{\mu }\partial_\mu  f^i(x,p)=\sum_{j=g,q,\overline{q}} C_{ij}(x,p)$
for the phase-space distribution function $f^i(x,p)$ of massless partons~\cite{Xu:2004mz} by employing elastic and radiative perturbative QCD cross sections.

\section{Evolution of azimuthal anisotropies}
\begin{figure}[h!]
	\centering
	\includegraphics[width=0.45\columnwidth]{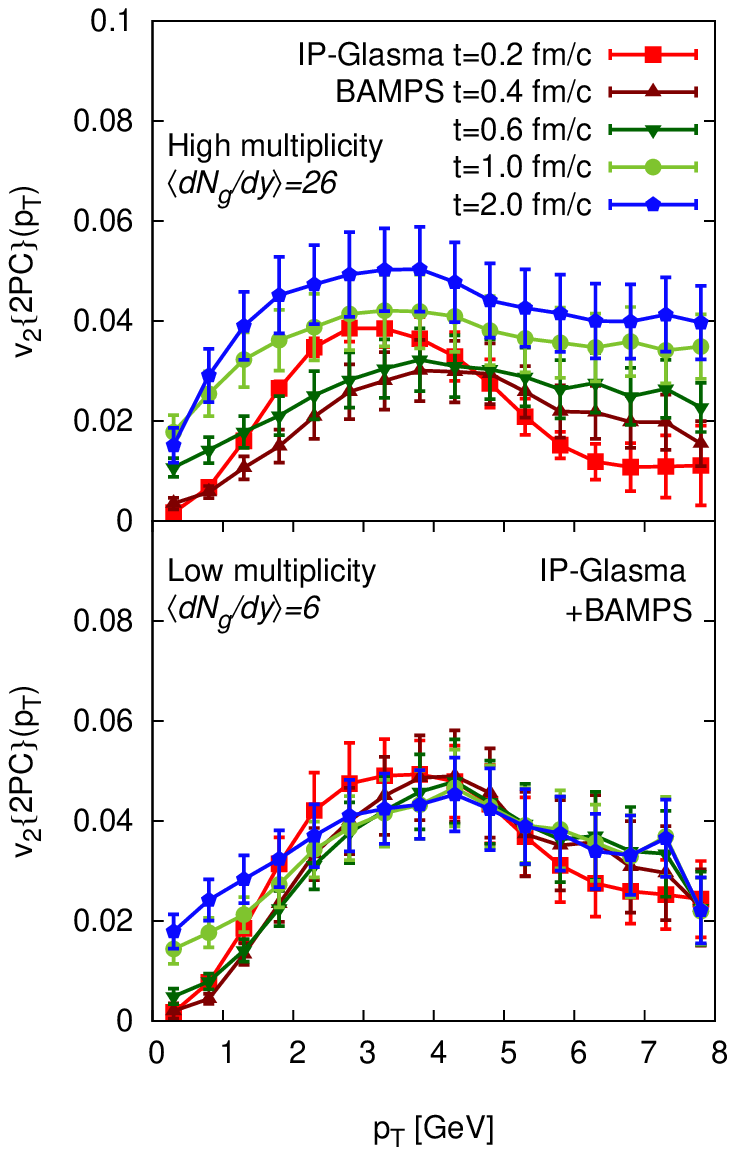}
	\includegraphics[width=0.45\columnwidth]{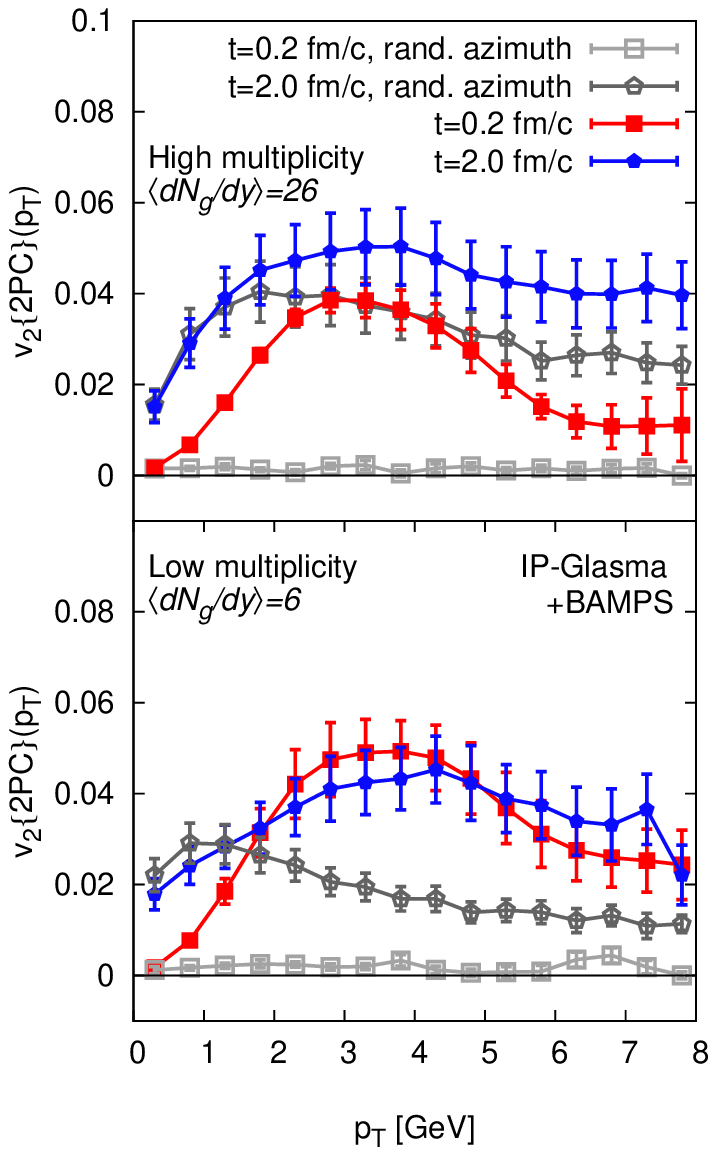}
	\caption{Left: Gluon $v_2\{2PC\}(\abspt)$ at mid-rapidity for different times in high multiplicity ($\langle dN_{g}/dy\rangle=26$, upper panel) and low multiplicity ($\langle dN_{g}/dy\rangle=6$, lower panel) p+Pb collisions. Right: Comparison of events including initial state momentum correlations (filled symbols) to the same events where the initial momenta were randomized in azimuth (rand. azimuth, open symbols).}
	\label{fig:v2_2PC}
\end{figure}

In the left panel of Fig.~\ref{fig:v2_2PC}, we show the evolution of the azimuthal momentum space anisotropy for $\sqrt{s}=5.02\,{\rm TeV}$ p+Pb collisions characterized by the Fourier harmonics $v_{n}\{2PC\}$ of the two-particle correlation function (following the experimental analysis method~\cite{Chatrchyan:2013nka}). We show $v_2\{2PC\}(p_T)$ at different times, $t=0.2\,\mathrm{fm/c}\,\mathrm{(initial)}-2\,\mathrm{fm}/c$ for low multiplicity $\left(0.5 < \left( \dd N_g/\dd y\right)/ \langle \dd N_g/\dd y \rangle  < 1\right)$ and high multiplicity $\left(\left( \dd N_g/\dd y\right)/ \langle \dd N_g/\dd y \rangle  > 2.5\right)$ events. While in both cases momentum correlations lead to a sizeable initial state $v_{2}$, the subsequent dynamics is quite different: In high multiplicity events, we observe a pronounced effect of the final state interactions such that the high initial anisotropy at intermediate momenta $(\abspt\sim2-5~\mathrm{GeV})$ is significantly reduced within the first $0.2~\mathrm{fm/c}$ evolution in the parton cascade, while at the same time the correlation strength at higher and lower momenta begins to increase. Subsequently, the azimuthal anisotropy increases for all $\abspt$ and as a result, the pronounced peak at around $\abspt\sim 3~\mathrm{GeV}$, present after the IP-Glasma stage, is washed out by the final state interactions. For low multiplicity events, in contrast, modifications due to final state effects are less significant, as the final curve $v_{2}(p_T)$ closely resembles that of the IP-Glasma initial state. Only at low transverse momenta the azimuthal anisotropy increases. These results confirm the basic expectation that final state effects gain significance as the density of the medium increases~\cite{Schlichting:2016xmj,Schlichting:2016sqo}.

 We find that the average number of interactions in low-multiplicity events ($N_{\rm{scat}}=4.5 \pm 1.1$) is almost the same as in high-multiplicity events ($N_{\rm{scat}}=5.6 \pm 1.1$), but the average number of \emph{large angle scatterings} is in fact significantly larger in high-multiplicity events $(N_{\rm{scat}}^{\rm{large~angle}}=1\pm 0.18)$ as compared to low-multiplicity events $(N_{\rm{scat}}^{\rm{large~angle}}=0.53 \pm 0.14)$.
In order to further disentangle the effects of initial state momentum correlations and final state response to geometry, we performed an additional set of simulations (labeled rand. azimuth) where the azimuthal angle of the transverse momentum $\pt$ of each gluon is randomized ($0<\varphi_\abspt<2\pi$) before the evolution in the parton cascade. Our results are compactly summarized in the right panel of Fig.~\ref{fig:v2_2PC}, where we compare $v_2\{2PC\}(p_T)$ in the different scenarios. No initial state momentum correlations are present in the rand. azimuth case (open gray symbols) and the initial state $v_2$ vanishes identically at $t=0.2~\mathrm{fm/c}$. However, over the course of the kinetic evolution a $v_{2}(p_T)$ of $\sim4\%$ at $\abspt\sim2~\mathrm{GeV}$ in high multiplicity events and $\lesssim 3\%$ at $\abspt~\sim 1~\mathrm{GeV}$ in low multiplicity events is built up. Nevertheless, for momenta above $\abspt\sim 2.0~\mathrm{GeV}$ (low multiplicity) and $\abspt\sim 4.0~\mathrm{GeV}$ (high multiplicity), the purely final state $v_2$ in the rand. azimuth scenario remains significantly below the initial state + final state $v_{2}$ of the full calculation, indicating the importance of initial state momentum correlations. 
\section{Initial state vs. final state effects}
\begin{figure}
	\centering
	\includegraphics[width=0.61\columnwidth]{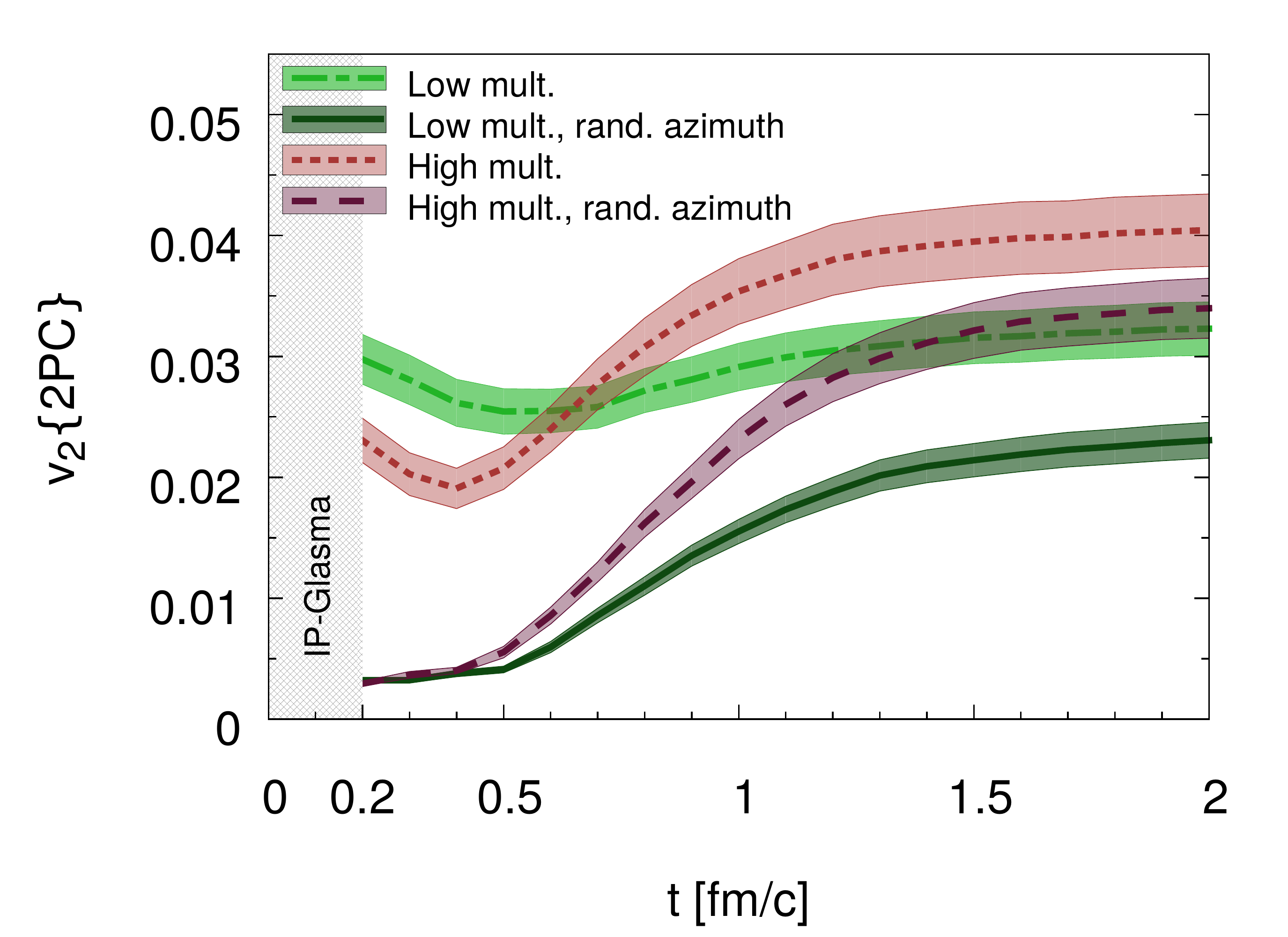}
	\caption{Evolution of the $\abspt$ integrated azimuthal anisotropy $v_{2}\{2PC\}$ for high and low multiplicity p+Pb events.}
	\label{fig:integratedV2}
\end{figure}

In Fig.~\ref{fig:integratedV2} we study the time-evolution of the $\abspt$ integrated $v_2\{2PC\}$. While in the rand. azimuth case the $v_2\{2PC\}$ is built up slowly as a function of time in response to the initial state geometry, a qualitatively different behavior emerges in the more realistic case including initial state correlations. In this case, large angle scatterings at early times begin to destroy initial state momentum correlations leading to an initial decrease of $v_2\{2PC\}$ as a function of time. This happens because the directions of the initial state anisotropy and the eccentricity responsible for generating the final state $v_2$ are generally uncorrelated. Subsequently, between $t\sim 0.5 - 1~\mathrm{fm/c}$ the response to the initial state geometry sets in, leading again to an increase of $v_2\{2PC\}$. 
We observe that the effect of initial state momentum correlations remains even in high-multiplicity events, where about $15\%$ of the flow can be attributed to the initial state, compared to about $45\%$ for low-multiplicity events.
\section{Conclusions}
\label{sec:Conclusions}
The observed large momentum anisotropies in small systems have challenged
our understanding of the space-time evolution of high-energy nuclear collisions. We developed a new framework including both initial state momentum correlations and final state interactions. Based on a weak-coupling picture of the space-time dynamics, we are able to consistently describe the gradual change from an initial state dominated to a final state dominated scenario, and find, that both initial state and final state effects can be quantitatively important for two-particle correlations in p+Pb collisions, depending on multiplicity. 

This work was supported by the Helmholtz International Center
for FAIR within the framework of the LOEWE program
launched by the State of Hesse. 
M.G., C.G. and S.S. acknowledge support by the
Deutsche Forschungsgemeinschaft (DFG) through the grant
CRC-TR 211 ``Strong-interaction matter under extreme conditions''.
S. S. acknowledges support
by the U.S. Department of Energy (DOE) under Grant
No. DE-FG02-97ER41014. B. P. S. is supported under
DOE Contract No. DE-SC0012704. Z. X. was supported
by the National Natural Science Foundation of China under
Grants No. 11575092 and No. 11335005, and the Major
State Basic Research Development Program in China under
Grants No. 2014CB845400 and No. 2015CB856903.
Numerical calculations used the resources of the Center
for Scientific Computing (CSC) Frankfurt and the National
Energy Research Scientific Computing Center, a DOE
Office of Science User Facility supported by the Office
of Science of the U.S. Department of Energy under
Contract No. DE-AC02-05CH11231.

\bibliographystyle{elsarticle-num}
\bibliography{library_manuell.bib}







\end{document}